\newcommand{\rbar}{$R_{\rm b}$}

\newcommand{\sbar}{$S_{\rm b}$}

\documentclass{aa}  

\usepackage{graphicx,natbib}
\usepackage{txfonts}
\usepackage[dvipsnames]{xcolor}

\begin{document}

\title{Properties of Barred Galaxies with the Environment: I. the case of the Virgo cluster }


\author{J. Alfonso L. Aguerri 
          \inst{1,2}, Virginia Cuomo \inst{3}, Azahara Rojas-Roncero\inst{2,4} \and Lorenzo Morelli \inst{3}
          }

   \institute{Instituto de Astrof\'{\i}sica de Canarias. C/ V\'{\i}a L\'actea s/n, 38200 La Laguna, Tenerife, Spain\\
              \email{jalfonso@iac.es}
         \and Departamento de Astrof\'{\i}sica de la Universidad de La Laguna, E-38206 La Laguna, Spain
        \and Instituto de Astronom\'{\i}a y Ciencias Planetarias, Universidad de Atacama, Avenida Copayapu 485, Copiap\'o, Chile
        \and Institut d'Estudis Espacials de Catalunya (IEEC), C/ Gran Capità 2-4, 08034 Barcelona, Spain}

   \date{\today}

 
  \abstract
   {Barred structures are widespread in a considerable fraction of galactic discs, spanning diverse environments and galaxy luminosities. The environment likely exerts a significant influence on bar formation, with tidal interactions leading to the emergence of elongated features resembling bars within galaxy discs. It is plausible that the structural parameters of bars resulting from tidal interactions in high-density galactic environments differ from those formed through internal disc instabilities in isolated galaxies. To empirically test this scenario, a viable approach is to compare the structural parameters of bars in galaxies situated within distinct environments. }
   {The objective of this study is to study environmental effects on the properties of bars by conducting a comparison between the two key structural parameters of bars, namely strength and radius, in galaxies situated within the Virgo cluster and galaxies of comparable luminosities found in environments characterized by lower galaxy densities.}
   {We have collected data on the bar radius and bar strength for a sample of 36 SB0 and SBa galaxies located within the Virgo cluster. These galaxies exhibit a large range of magnitudes, with values ranging from $M_{r} = -22$ to $M_{r} = -17$. Additionally, we analyzed a sample of 46 field galaxies with similar morphologies and luminosity range. The measurements of bar parameters were conducted by employing Fourier decomposition on the $r$-band photometric images of the galaxies.}
   {The analysis reveals that the bar radius exhibits a correlation with galaxy luminosity, indicating that larger bars are typically found in more luminous galaxies. When comparing galaxies with fixed luminosities, the field galaxies display larger bar radii compared to those in the Virgo cluster. However, when the bar radius is scaled by the size of the galaxy, the disparity diminishes and the scaled bars in the Virgo cluster and the field exhibit similar sizes. This is because galaxies of similar luminosities tend to be larger in the field environment compared to the cluster and because the bars adapt to the discs in which they live. Regarding the bar strength, no significant differences were observed for bright galaxies ($M_{r} < -19.5$) between those located in the Virgo cluster and those in the field. In contrast, faint galaxies ($M_{r} > -19.5$) show stronger bars in the field than in the cluster.}
   {The findings of this study indicate that the size of galaxies is the parameter that is influenced by the environment, while the bar radius remains independent of the environment when scaled by the galaxy size. The findings of this study indicate that the environment influences the size of galaxies rather than the bar radius, which remains independent of the environment when scaled by the galaxy size. Regarding the bar strength there is no influence of the environment for bright galaxies. However, bars in faint galaxies are weaker in the cluster environment. This could be explained by an enhancement of disc thickness in dense environments which is more efficient in faint galaxies. These results support the notion that the internal dynamics and intrinsic characteristics of galaxies play a dominant role in the formation and evolution of bars, regardless of the surrounding environment.}

   \keywords{galaxies: structure --
                galaxies: evolution
               }
               
\authorrunning{J. A. L. Aguerri et al.}
\titlerunning{Properties of Bars in the Virgo Cluster}
   \maketitle
%

\section{Introduction}

Bars are elongated stellar structures found in a large fraction of galaxies in both the nearby \citep[][]{Aguerri2009} and distant Universe \citep{Guo2023}. The presence of a bar in a disc galaxy strongly modifies its dynamics by exchanging angular momentum between different structures within the galaxy \citep[][]{Debattista2000, Athanassoula2002, Athanassoula2013}.

The formation of a bar is a complex process that depends on a large number of internal and/or external properties of the galaxy \citep[see, e.g.,][]{Athanassoula2013}. So far, three main scenarios for bar formation have been proposed. In the first scenario, bars are formed due to internal galactic properties, such as the gas fraction or the shape of the dark matter halo. In this case, the disc becomes unstable and forms a bar in a timescale of about 1-1.5 Gyr. The bar can undergo buckling instabilities \citep{Raha1991}, temporarily weakening it, before it grows continuously by secular evolution over several Gyr, by exchanging angular momentum between the disc and other structures within the galaxy \citep{Athanassoula2013}. In the second scenario, bars can be formed by tidal interactions with another galaxy. These interactions produce elongated bar-like features in the discs of the galaxies. In this case, the bar formation process is different from the internal process, and bars form over longer timescales and grow slowly due to the limited exchange of angular momentum between the disc and other components of the galaxy \citep[see][]{MartinezValpuesta2016}. A third scenario for bar formation has been recently proposed by \cite{Yoon2019}, suggesting that cluster-cluster interactions can be responsible for the formation of these structures in galaxies.

The environment could influence the properties of the bars in at least two of the proposed bar formation scenarios. For this reason, we propose to analyze the influence of the environment on the properties of a sample of bars located in one of the most massive structures in the nearby Universe: the Virgo cluster. The effect of the environment on bar formation can be analyzed using two different approaches. The first approach is to study the relation between the bar fraction and the environment where the galaxies are located. The second approach is to analyze the properties of bars, such as bar radius ($R_{b}$), strength ($S_{b}$), and pattern speed ($\Omega_{b}$), in different environments.


The impact of the environment on bar formation has been extensively discussed in the literature. Several studies have found a clear correlation between the bar fraction and the environment of galaxies, indicating that bars are more frequent in high-density environments \citep[see][for example]{thompson1981, eskridge2000, barway2011, skibba2012, lin2014}. However, other investigations have failed to identify a significant relationship between the bar fraction and the environment \citep[see][for example]{vandenbergh2002, li2009, barazza2008, Aguerri2009, Cameron2010, MendezAbreu2010, marinova2012, smith2022}.

The relationship between the bar fraction and the environment might also be influenced by other galaxy properties such as their luminosity or morphology. The distribution of the bar fraction as a function of galaxy luminosity varies significantly from cluster to field environment. For instance, \cite{MendezAbreu2012} studied the bar fraction in three different environments ranging from the field to the Virgo and Coma clusters and found a large difference between the bar fraction distributions as a function of galaxy luminosity in the field and Coma Cluster, with Virgo being an intermediate case. Barred galaxies peaked at $M_r\simeq -20.5$ mag in clusters and at $M_r\simeq-19.0$ mag in the field. This was interpreted as a variation in the effect of the environment on bar formation depending on galaxy luminosity: brighter disc galaxies are stable enough against close interactions to maintain their discs cold, so the bar formation is triggered by interactions when the galaxies are probably in a pre-cluster stage. For fainter galaxies, interactions become strong enough to heat up the discs, inhibiting bar formation. This trend has been recently confirmed up to $z\sim0.4$, using data from the JWST telescope \citep{MendezAbreu2023}.

Similar conclusions were drawn by \cite{lin2014}, who analyzed $\sim30,000$ barred galaxies in the local Universe and their environment with Sloan Digital Sky Survey (SDSS) data. After removing any dependence on stellar mass, color, and stellar surface mass density, they found that the clustering of barred and unbarred galaxies is different when splitting the sample into early- and late-type galaxies. In fact, early-type barred galaxies seem to be more strongly clustered on scales from a few hundred kpc to 1 Mpc when compared to early-type unbarred galaxies. At these intermediate scales, the correlation function is dominated by the one-halo term, which would indicate that barred early-type galaxies are more frequently satellite systems. This is similar to what \cite{barway2011} found in S0 galaxies: a higher bar fraction in clusters rather than in the field. Moreover, barred late-type galaxies have few neighbors within $\sim50$~kpc since tidal forces from close companions suppress the formation/growth of bars.

In the study by \cite{Tawfeek2022}, the distribution of barred galaxies was analyzed across 32 galaxy clusters, revealing that bars were present in approximately $\sim 30\%$ of the analyzed galaxies when studied in the optical bands. The study found that the bar fraction exhibited a dependence on both galaxy mass and morphological type, being highest for massive late-type galaxies. However, the fraction of barred galaxies decreased with increasing cluster mass and decreasing clustercentric distance. When accounting for morphological type, the higher fraction of barred galaxies was consistently found among the latest morphological types, regardless of their location within the cluster. This suggests that the presence of bars is driven by the galaxy morphological transformation associated with in-fall toward the cluster, leading to the formation of dynamically hot systems with early-type morphology. At larger clustercentric distances, the bar fraction increased with distance to the nearest neighbor galaxy, suggesting that bars could be suppressed or even destroyed by the presence of a companion. The sample of analyzed barred galaxies consisted of either early-type, star-forming galaxies located within the virial radii of the clusters, or late-type quenched galaxies located in the outer regions of the clusters.

In contrast, \cite{castignani2022} introduced the effect of the cosmic web on the bar fraction, analyzing the bar fraction in various environments, including clusters, filaments, and fields. The study observed a slight decrease in the bar fraction from high-density environments such as clusters to filaments and fields.


The structure of bars in galaxies can be described by three main parameters: their radius, strength, and pattern speed. The radius and strength of a bar are related to the prominence of the non-axisymmetric component of the galactic disc potential, with larger and stronger bars indicating more significant changes in the disc structure. The pattern speed of the bar determines its dynamics and the location of the bar resonances, which play a crucial role in the formation of different orbit families. However, the relationship between these bar parameters and the galactic environment has received less attention in the literature.

In their study, \cite{smith2022} investigated how the environment affects various properties of galaxies, including spiral arms, bars, concentration, and quenching. They observed that for field galaxies, there is no significant correlation between bar strength and stellar mass. However, they did find a weak correlation between the two for cluster galaxies. Additionally, the study revealed a strong correlation between galaxy concentration, which refers to the prominence of the bulge, and bar strength in both field and cluster galaxies. This result contradicts the idea that there is an excess population of tidally-induced bars in clusters due to the expected enhancement of bar strengths by tidal forces from the cluster as a whole \citep{Lokas2020}. The researchers also examined the specific star formation rate (sSFR), which is the star formation rate per unit of stellar mass, and its relationship with bar strength. No clear correlation was found for either field or cluster galaxies.

Additionally, \cite{Marinova2010} studied the properties of barred galaxies in the central region of the Coma cluster, the densest environment in the nearby Universe. They focused on disc galaxies, which are dominated by lenticular ones, and they identified an optical bar fraction of almost $50\%$. The bars ended up being relatively small (with $R_{\rm b}<2$~kpc) and not particularly strong, as observed in S0 galaxies within less dense environments.

\cite{Erwin2012} studied the surface-brightness profiles of discs in S0 galaxies in the local field and those in the Virgo Cluster. While in the field, discs present the three main types (Types I, II, and III: single-exponential, truncated, and antitruncated), in Virgo truncated discs are not present while single-exponential ones are significantly more common. This could be due to environmental and/or bar effects. Indeed, \cite{Erwin2008,Erwin2012} suggested that truncations are predominantly related to the bar-OLR interactions, which in turn are strengthened by the presence of significant gas in the outer disc. S0s in the field may be able to retain gas in their outer disc long enough to show the effects of bar-OLR interactions and develop Type II profiles. On the other hand, S0 galaxies in Virgo may have lost their gas, particularly in the outer disc, due to a combination of ram-pressure stripping and strangulation, which explain the observed lack of truncated discs. Additionally, the analysis of the structural parameters of the discs of galaxies located in high-density environments shows differences with similar discs in the field. In particular, the scale length of the discs of galaxies in the Coma cluster was found to be smaller than that of similar galaxies in the field \citep[][]{aguerri2004, gutierrez2004}.


\cite{lokas2016} analyzed the formation of a tidal-induced bar in a Milky Way-like disc located in the Virgo cluster. In all simulations, tidally-induced bars were formed for all studied orbits after the first pericenter passage. The formed bars were longer, stronger, and slower than those formed in isolation. These bars were formed by transferring angular momentum from their stellar component to their dark matter halos. Few strippings of the stellar component were accounted for in these models. The simulations proposed by \cite{Aguerri2009b} formed bars by tidal interactions in galaxies losing a large fraction of their total and stellar mass. In all models, bars appeared in the remnants of the discs. These bars showed large bar radius ($R_{b}$) compared with the scale lengths ($h$) of the discs ($R_{b}/h \approx 2$). Other simulations analyzing fast close galaxy encounters did not find differences in the structural parameters of the bars before and after the interaction occurred \citep{2017MNRAS.464.1502M}. In some cases, no bars were formed by interactions \citep[see][]{gnedin2003,smith2015}.

This work focuses on analyzing the properties of bars ($R_{\rm b}$ and $S_{\rm b}$) in a sample of barred galaxies located in the Virgo cluster. We have selected a range of barred galaxies with different luminosities in the cluster and compared their bar radii and strengths with galaxies of similar luminosities located in the field. The paper is organized as follows: in Section 2, we present the selected galaxies. In Section 3, we describe the measurement of the bar properties and compare galaxies in different environments. Finally, Sections 4 and 5 present the discussion and conclusions, respectively. The cosmology adopted for this work is $H_0 = 70$ km s$^{-1}$ Mpc$^{-1}$, $\Omega_m = 0.3$, and $\Omega_\Lambda = 0.7$.
\section{Sample of Barred Galaxies}

Two samples of barred galaxies were chosen for this study to investigate the impact of the environment on the structural characteristics of bars. One sample included barred galaxies located within the Virgo cluster, while the other sample included barred galaxies not located in a cluster environment. Both samples were limited to galaxies with SB0 and SBa morphological types, to avoid possible morphological differences, and with intermediate disc inclinations ($25\degr<i<75\degr$) to facilitate the necessary isophotal and Fourier analysis required for measuring the bar parameters.

\subsection{Barred galaxies from Virgo}

The Virgo cluster is the closest massive structure to us, with a total mass of $1.2 \times 10^{15} M_{\odot}$ and a virial radius of $R_{vir} = 1.72$ Mpc \citep[][]{hoffman1980}. This cluster contains a diverse range of galaxy types and environments due to the fact that it is not yet virialized. Indeed, several groups of galaxies are falling into the cluster \citep{Kashibadze2020}. This makes the Virgo cluster a suitable environment for studying galaxy evolution. Its proximity \citep[at a distance of 18.7 Mpc,][]{Davies2014} allows us to achieve a spatial resolution such that 1 arcsecond on the sky corresponds to 0.09 kpc. This spatial resolution provides an opportunity to resolve structures as small as a few kpcs, such as bars in galaxies located in the Virgo cluster. The sample of barred galaxies in Virgo was selected using information from two catalogs of galaxies: the Virgo Cluster Catalog (VCC) and the Extended Virgo Cluster Catalog (EVCC).

The VCC was first published by \cite{Binggeli1985}, and it includes 2096 galaxies distributed across an area of approximately 140 deg$^{2}$ around the center of the Virgo cluster, which is located at the position of the central galaxy M87. The catalog provides information on visual morphological classification and cluster membership, among other parameters. The EVCC, an extension of the VCC, was created by \cite{Kim2014} using additional data from the Sloan Digital Survey Data Release 7 \citep[SDSS-DR7][]{Abazajian2009}. This catalog covers a larger area of approximately 725 deg$^{2}$ around the cluster center and includes 1589 galaxies.

\begin{figure}
   \centering
\includegraphics[width=9cm]{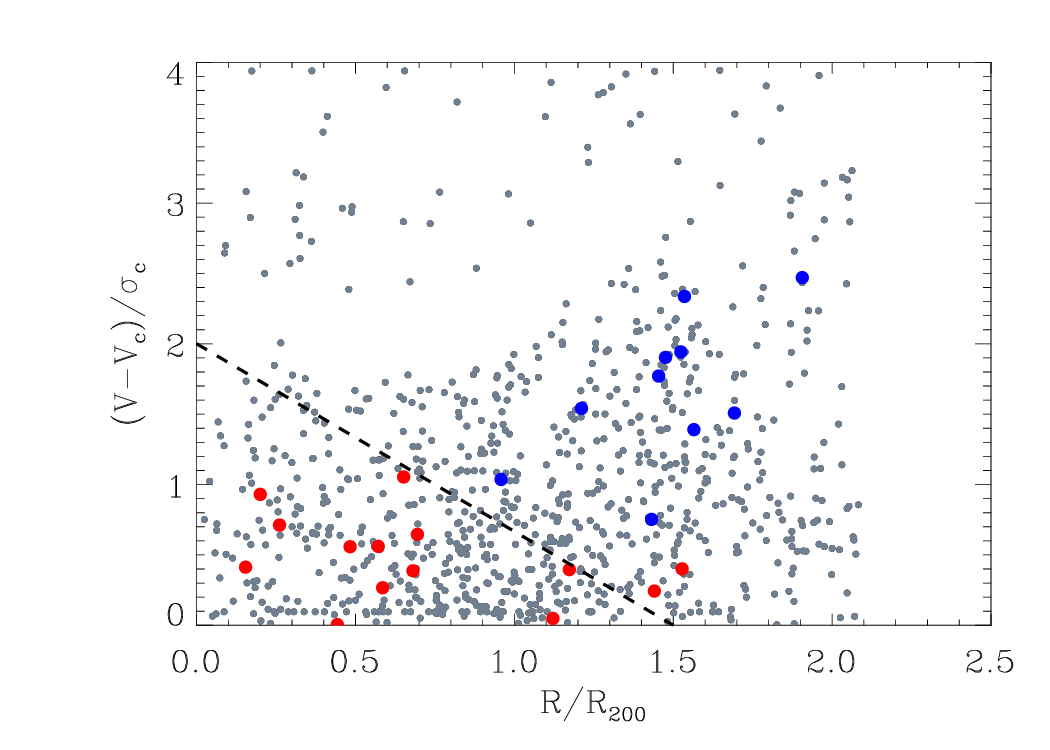}
\caption{Distribution of the barred galaxies catalogued as cluster members (red points) and possible cluster members (blue points), for which recession velocity is available. The gray points represent other galaxies in the VCC and EVCC catalogs. The diagonal dashed line separates the virialized (bellow the line) and the infalling (above the line) cluster regions \citep[see][]{oman2013}.}
    \label{caustic}%
\end{figure}

A total of 174 galaxies were morphologically classified as barred (SB) in the VCC and EVCC catalogs, with 48 and 126 from VCC and EVCC, respectively. The number early-type barred galaxies (SB0 or SBa types) reduces to a final sample of 60 galaxies. The catalogs provided information on the cluster membership classification of galaxies, with classifications as cluster members, possible cluster members, or background objects based on their recession velocity or morphological features. The composition of the sample according to their morphological type and cluster membership is shown in Tab. \ref{tab:catalogs}. Figure \ref{caustic} displays the distribution of the galaxies in the phase-space diagram for those with radial velocity reported. Barred galaxies classified as cluster members (red points) are located inside the virialized cluster region, while those cataloged as possible cluster members (blue points) are located in the infalling region. In this study, galaxies cataloged as possible cluster members were considered members, resulting in a total of 36 barred galaxies as cluster members.

\begin{table}[]
    \centering
    \caption{Number of barred SB0 and SBa galaxies catalogated as members, possible members and background galaxies in the VCC and EVCC catalogs.}
    \begin{tabular}{cccc}
    \hline
    & Members & Possible members & Background \\
    \hline
    SB0 & 11 & 3 & 8 \\
    SBa & 14 & 8 &16 \\
    \hline
    \end{tabular}
    \label{tab:catalogs}
\end{table}

Total $r$-band magnitudes were obtained for all selected galaxies and converted to absolute $r$-band magnitudes $M_r$, using the galaxy distance reported by NED\footnote{The NASA/IPAC Extragalactic Database is available at \url{https://ned.ipac.caltech.edu/}} based on the radial velocity with respect to the cosmic microwave background reference frame and assuming a value of $H_0=70$~ km~s$^{-1}$~Mpc$^{-1}$. The sample of barred galaxies in the Virgo cluster covers a range of absolute magnitudes in the $r$-band of [-21.3, -16.0].

\subsection{Barred galaxies from field}

In order to compare bar properties of galaxies in different environments, we compiled a sample of barred galaxies from the field. To ensure a comparable range in absolute magnitude, we gathered results from various analyses.

Firstly, we collected the SB0 and SBa galaxies from the VCC and EVCC catalogues classified as background objects. This sample of field galaxies is formed by 8 SB0 and 16 SBa galaxies (see Tab. \ref{tab:catalogs}). This sample of 24 objects span a range of absolute magnitude in the $r$-band of [-19.5, -18.0]. 

Secondly, we collected barred galaxies from the literature to complete the bright part of the sample. Two set of barred galaxies from the literature were considered. The first one is form by barred SB0 and SBa galaxies with measured bar parameters included in \cite{corsini2011}, for which SDSS images are available. These galaxies were selected because the corresponding bar properties were already available, being measured using techniques comparable to those used here. This includes 7 SB0 and 3 SBa galaxies. Finally, we revised the galaxies from Calar Alto Legacy Integral Field Area survey \citep[CALIFA,][]{Sanchez2012}, whose aim was to measure the properties of a statistically significant sample of nearby galaxies with integral field spectroscopy. From 
the CALIFA mother-sample, which includes $\sim950$ objects with available SDSS images, we selected all the strongly barred galaxies, which were visually classified as SB by \cite{Walcher2014}. Using both the morphological classification from the survey and from NED, we selected all the SB0 and SBa included in the CALIFA mother-sample, corresponding to 102 galaxies. We adopted disc properties (position angle PA and inclination $i$) from the photometric-decomposition on the SDSS $r$-band images presented by \cite{MendezAbreu2017}, when available, or performing an isophotal analysis in the disc region following the prescriptions of \cite{Cuomo2019} and \cite{Buttitta2022}, after retrieving the SDSS $r$-band images from the science archive of the DR17 \citep{Abdurro'uf2022}. We ended up with 10 SB0 and 12 SBa galaxies, after identifying galaxies with the selected inclination range.

The total sample of field galaxies includes 32 objects, with 17 SB0 and 15 SBa galaxies. Total $r$-band magnitude were retrieved for all the selected galaxies, and translated into absolute $r$-band magnitude $M_r$. The sample of field galaxies covers a range [-22.6,-18.0] in absolute $r$-band magnitude. 

\subsection{Galaxy Densities for the cluster and field samples}

We calculated the galaxy density around all galaxies in both the cluster and field samples using data from SDSS-DR17. To obtain this density, we downloaded a catalog from SDSS-DR17 containing galaxies located within a 10-deg radius from the Virgo cluster center, assumed to be at the position of M87. Within this catalog, we considered objects with radial velocity within $v_{c} \pm 3 \sigma_{c}$ as Virgo galaxies, where $v_{c}$ and $\sigma_{c}$ are the Virgo cluster velocity and velocity dispersion, respectively. For the barred galaxies in the field sample, we downloaded catalogs from SDSS-DR17 containing galaxies located within 500 kpc around each galaxy in the sample. The galaxy density $(\Sigma_{g})$ around each galaxy was measured by taking into account the number of galaxies with $r$-band magnitude $m_{r} > 18.0$ located within a radius of 500 kpc and within $\pm 1500$ km s$^{-1}$ around each galaxy in the sample. We adopted the magnitude limit of $m_{r}=18.0$ given the SDSS spectroscopic magnitude limit. Furthermore, we corrected the galaxy density by the spectroscopic completeness of the SDSS data.

\begin{figure}
   \centering
   \includegraphics[width=9cm]{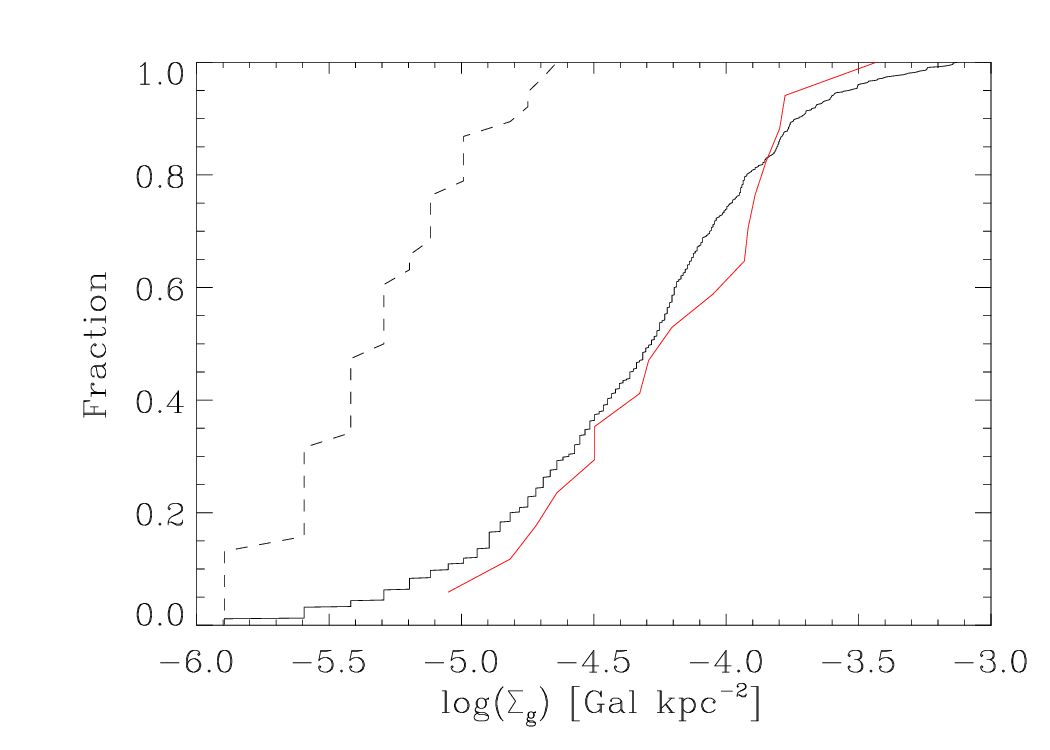}
   \caption{Local environment of the galaxies in the Virgo cluster (full black line), barred member galaxies (red line), and field barred galaxies (dashed black line).}
\label{density}%
    \end{figure}

Figure \ref{density} shows the cumulative distribution functions of the galaxy density for galaxies in the Virgo cluster, barred galaxies from this study in both the cluster and field samples. The plot clearly shows that the entire sample of galaxies from the Virgo cluster and the subsample of barred ones have similar cumulative distribution functions, indicating that barred galaxies are equally distributed as non-barred ones within the Virgo cluster. In contrast, the sample of field galaxies used in this study is located in regions with lower galaxy densities.

Note that the galaxy sample from the field is more diverse compared to the sample of barred galaxies in the Virgo cluster. Our intention was to acquire two sets of galaxies covering the same range of luminosities and situated in distinct galactic density surroundings. Nevertheless, we lack information concerning whether the field galaxies are located within groups or are isolated.

\section{Results}

\subsection{Bar structural parameters: radius and strength}

\begin{figure}
    \centering
    \includegraphics[scale=0.51]{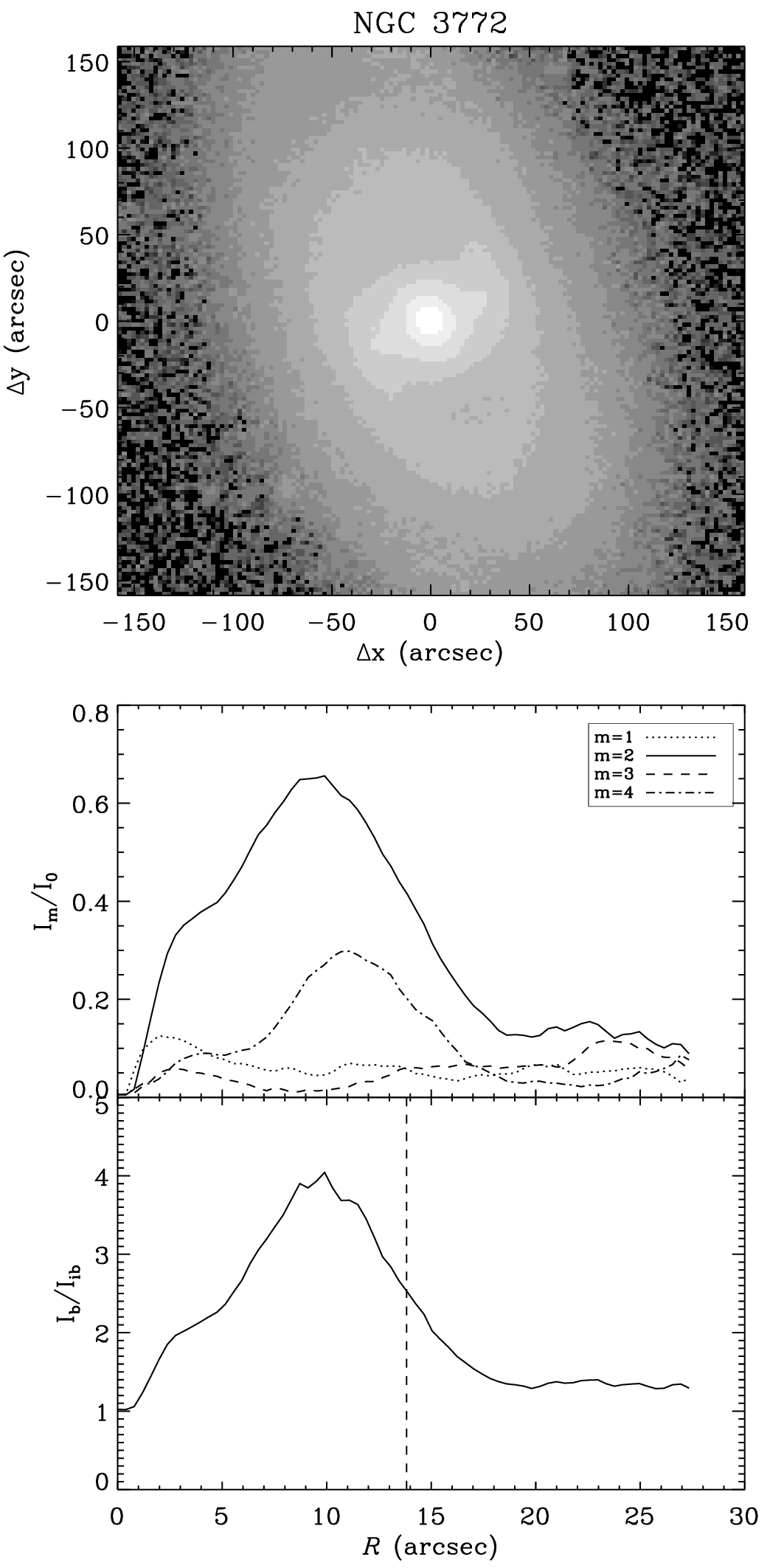}
    \caption{Fourier analysis for NGC~3772, taken as example. From top to bottom: non-deprojected SDSS $r$-band image of the galaxy, radial profiles of the relative amplitude of the $m =1, 2, 3,$ and 4 Fourier components, radial profiles of the bar/interbar intensity ratio used to derive the bar radius, marked by the vertical dashed line.}
    \label{fig:fourier}
\end{figure}

Fourier analysis has been largely used to detect different galactic structures, and it is particularly suited to characterise bars, which correspond to bisymmetric departures from axisymmetry \citep{Ohta1990,athanassoula2003,Cuomo2019b}. In this work we derived the photometric properties of bars applying a Fourier analysis on the azimuthal light distribution of the galaxy \citep{Aguerri2000} derived from the SDSS $r$-band images. The analysis was applied to the sample of barred galaxies from Virgo and to those from the field for which a similar analysis was not available from the literature. Figure~\ref{fig:fourier} shows an example of the Fourier analysis performed here.

First, we deprojected the images of the galaxy by stretching the original ones along the disc minor axis by a factor equal to $1/\cos i$, and conserving the flux. 

Then, we decomposed the deprojected azimuthal surface brightness profile $I(r,\phi)$ of each sample galaxy, assuming polar coordinates in the galaxy disc $(r,\phi)$ into a Fourier series

\begin{equation}
    I(r, \phi) = \frac{A_0 (r)}{2} + \sum A_m (r) \times \cos(m\phi) + \sum B_m (r) \times \sin(m\phi)
\end{equation}

We derived the radial profiles of the amplitudes of the Fourier components $I_m(r)$ as

\begin{equation}
\begin{split}
    I_0 (r) & = A_0 (r)/2 \\
    I_m (r) & = (A_m^2(r) + B_m^2(r))^{1/2}
\end{split}
\end{equation}

for $m = 0, 1, 2,3,4,5,6$ and of the phase angle $\phi_2$ of the $m = 2$ Fourier component as was originally done by \cite{Aguerri2000} and following the prescriptions of \cite{Cuomo2019b,Buttitta2022}. Strongly barred galaxies present large even Fourier components, a clearly peaked $m=2$ one, and a constant behaviour of $\phi_2$ within the bar region (see Fig.~\ref{fig:fourier}).

The bar radius \rbar\ is derived from
the luminosity contrasts between the bar and interbar intensity as a
function of radial distance \citep{Aguerri2000}. Indeed, the even Fourier components define the bar profile as $I_{\rm b} (r) = I_0(r) + I_2(r) + I_4(r)+I_6(r)$, while the odd ones define the interbar one $I_{\rm ib} (r) = I_0 (r)-I_2 (r)+I_4 (r)-I_6 (r)$. Therefore, the length of the bar is calculated as the location where

\begin{equation}
    I_{\rm b}/I_{\rm ib} (R_{\rm b}) = \frac{(I_{\rm b}/I_{\rm ib})_{max} - (I_{\rm b}/I_{\rm ib})_{min}}{2} + (I_{\rm b}/I_{\rm ib})_{min}
\end{equation}

Selecting the bar radius based on Equation (3) is analogous to considering the full width at half maximum (FWHM) of the $I_{b}/I_{ib}$ curve. This method was later employed by \cite{Athanassoula2002} on analytic models, and its accuracy in measuring bar length was demonstrated. \cite{Ohta1990} previously opted for a fixed value of $I_{b}/I_{ib}=2$ for the bar radius, but this approach lacks precision across a wide range of bar luminosities found in galaxies \citep[see e.g.][]{Aguerri2000}.

The bar strength \sbar\ is calculated using the maximum of the ratio between the $m=2$ and $m=0$ Fourier components as

\begin{equation}
    S_{\rm b} = (I_2/I_0)_{max}.
\end{equation}

Per each galaxy from both the samples, we collected the Petrosian radii enclosing 50\% and 90\% of the galaxy light, $R_{\rm 50}$ and $R_{\rm 90}$, respectively, provided by the SDSS in the $r$-band. These radii can be used to give an estimation of the extension of the galaxies. In particular, we used them to obtain the relative length of the bars with respect to the size of the galaxies.

\begin{table}[]
    \centering
    \caption{Mean value and corresponding standard deviation of the bar parameters for the sample of Virgo and field galaxies.}
    \begin{tabular}{lcc}
    \hline
    Parameter & Virgo & field \\
    \hline
    $\langle R_{\rm b}\rangle$ [kpc] & $2.6\pm1.5$ & $6.1\pm3.0$ \\
    $\langle S_{\rm b}\rangle$ & $0.4\pm0.2$ & $0.5\pm0.2$ \\
    $\langle R_{\rm 50}\rangle$ [kpc] & $1.2\pm0.4$ & $3.3\pm1.1$ \\
    $\langle R_{\rm 90}\rangle$ [kpc] & $3.4\pm1.0$ & $9.3\pm4.0$ \\
    $\langle R_{\rm b}/R_{\rm 50}\rangle$ & $1.8\pm0.9$ & $1.8\pm0.7$\\
    $\langle R_{\rm b}/R_{\rm 90}\rangle$ & $0.6\pm0.3$ & $0.7\pm0.3$\\
    \hline
    \end{tabular}
    \label{tab:mean_value}
\end{table}

\subsection{Bar properties in the cluster and field environments}

\subsubsection{Bar radius as a function of the environment}

\begin{figure}
   \centering
\includegraphics[width=9cm]{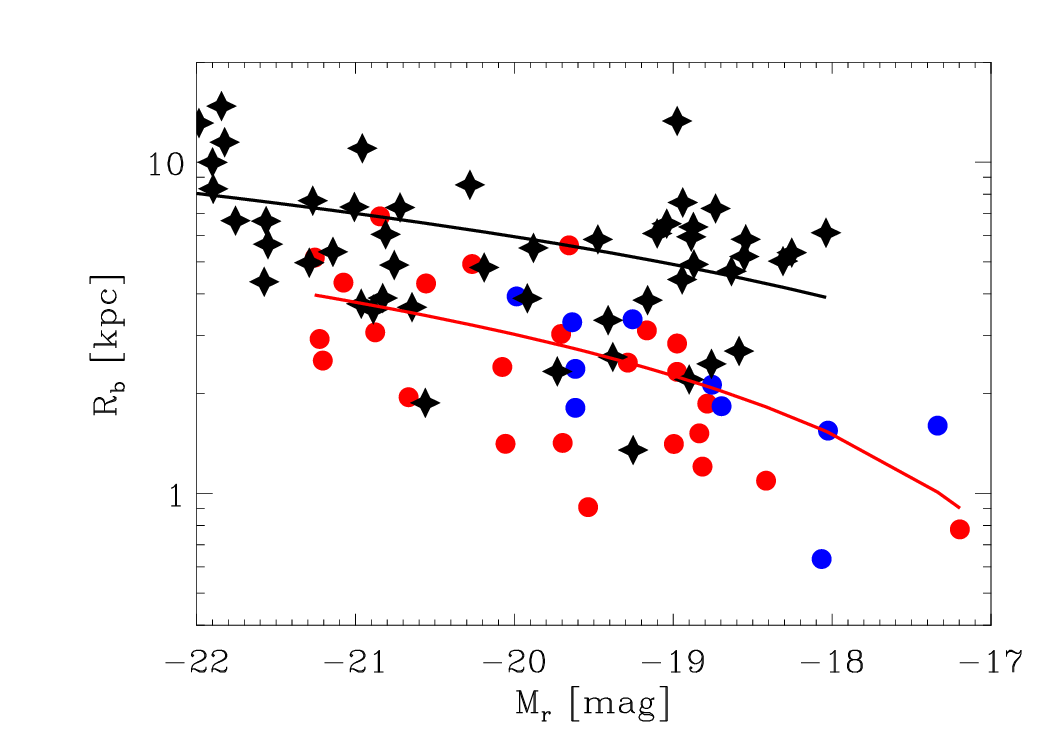}
\caption{Bar radius in kpc for the galaxies in the Virgo cluster (blue and red points) and from the field (black stars). The black and red full lines represent linear fits to the $R_{b}$ - $M_{r}$ relation for the field and Virgo galaxies, respectively.}
    \label{rbarkpc}%
\end{figure}

We initially compared the measured value of \rbar\ for the two samples of barred galaxies, namely, Virgo and field. Figure \ref{rbarkpc} illustrates the bar radius in kpc of both the field and Virgo galaxies as a function of their $r$-band absolute magnitude. The linear fits to the $R_{b}-M_{r}$ relations in our sample are $R_{b} = -12.08 - 0.75 M_{r}$ for the Virgo galaxies and $R_{b} = -14.92 - 1.04 M_{r}$ for the field galaxies. These linear relations demonstrate that the bar radius is dependent on the galaxy's luminosity, with more luminous galaxies exhibiting larger bars. This characteristic is consistent with both the field and Virgo barred galaxies and has been observed in other samples of barred galaxies with varying luminosities \citep[see][]{Cuomo2020}.

Furthermore, Figure \ref{rbarkpc} also indicates that bars in the Virgo cluster tend to be shorter than those in the field for a fixed value of $M_{r}$. Specifically, the mean values of $R_{b}$ for the Virgo and field galaxies are 2.6 and 6.1 kpc, respectively (see Tab. \ref{tab:mean_value}). This implies that the mean bar size in the field is approximately 3.5 kpc larger than that in the Virgo cluster, i.e. more than double. The presence of short bars in cluster has been also reported in the core regions of the Coma cluster. Thus, \cite{Marinova2010} shows that for a sample of 10 early-type bars in the core of the Coma clusters were always shorter than 2 kpc. This radius is smaller than the obtained for bars on similar galaxies out from clusters.

\begin{figure}
   \centering
   \includegraphics[width=9.5cm]{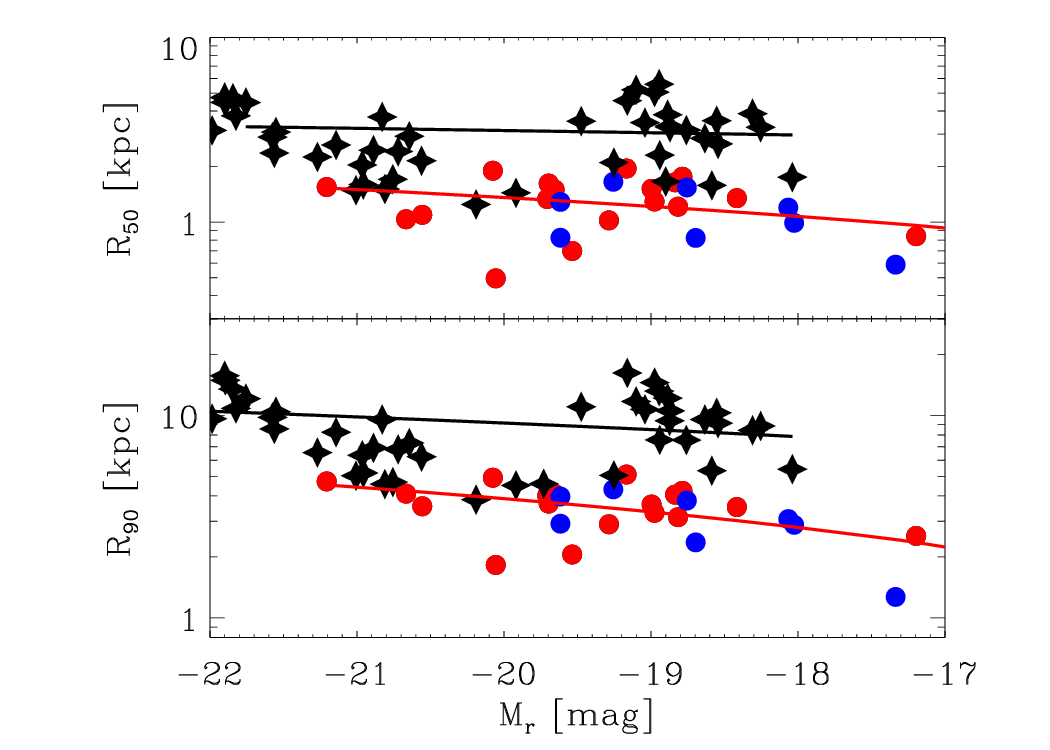}
   \caption{Galaxy radius enclosing 50$\%$ (top panel) and 90$\%$ (bottom panel) of the light in the $r$-band for the barred galaxies in the Virgo cluster (blue and red points) and in the field (black stars). The black and red full lines represent the linear fits to the Virgo and field galaxies, respectively.}
\label{r90lum}%
    \end{figure}

Figure \ref{r90lum} illustrates the relation between $R_{50}$ and $R_{90}$ with $M_{r}$ for the Virgo and field samples. We can see that more luminous galaxies are larger than less luminous ones. Additionally, it can be observed that for a fixed absolute magnitude, galaxies in the field show larger values of $R_{50}$ and $R_{90}$ than their counterparts in the cluster. The mean values of $R_{50}$ and $R_{90}$ for galaxies in Virgo and the field are given in Tab. \ref{tab:mean_value}. Note that the difference between Virgo and field galaxies is larger when considering the $R_{90}$ values rather than the $R_{50}$ ones. This indicates that the outermost parts of the galaxies in the Virgo cluster are more affected by the environment than the inner ones. The fact that barred galaxies in the field are larger than those in the Virgo cluster could explain the differences in the bar radius observed in Fig. \ref{rbarkpc}. 

\begin{figure}
   \centering
   \includegraphics[width=9.5cm]{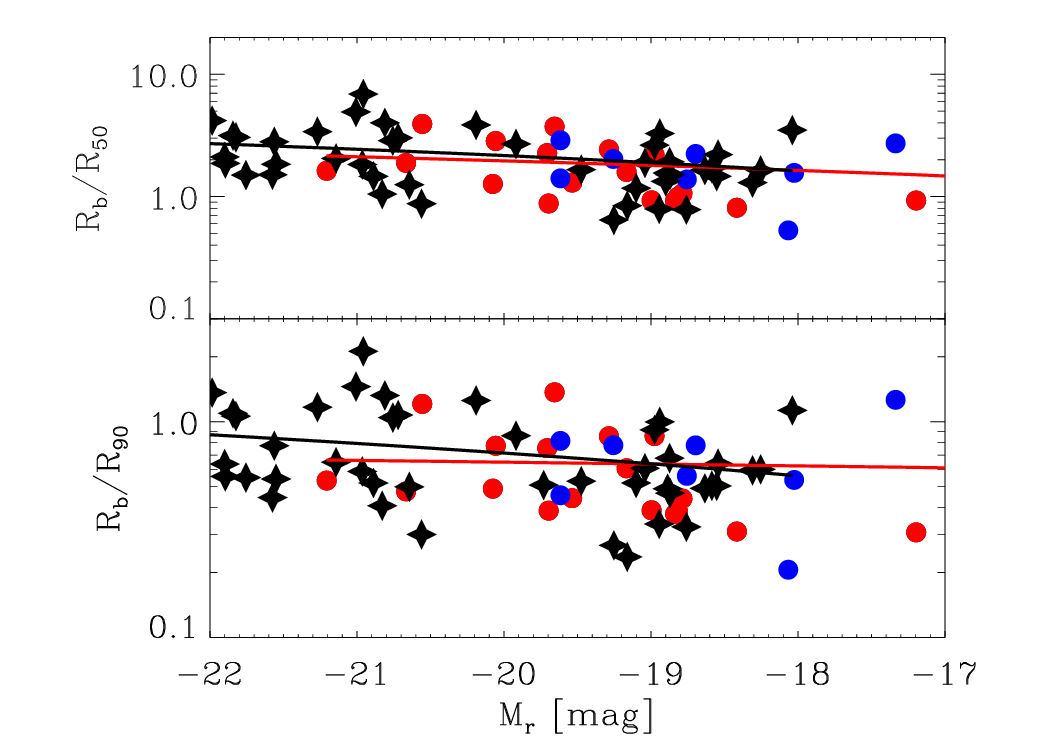}
   \caption{Bar radius scaled by $R_{\rm 50}$ (top panel) and $R_{\rm 90}$ (bottom panel) for the galaxies in the Virgo cluster (blue and red points) and in the field (black stars). The full black and red lines represent linear fits to Virgo cluster and in the field, respectively.}
              \label{rbarr90}%
    \end{figure}

Indeed, when we scaled the \rbar\ by the galaxy extension ($R_{50}$ or $R_{\rm 90}$), the Virgo's and the field's bar radii become similar (see Fig.\ref{rbarr90} and Tab.\ref{tab:mean_value}). This result implies that the environment does not play an important role in the radius of the bars.

\subsubsection{Bar strength as a function of the environment}

\begin{figure}
   \centering
   \includegraphics[width=9cm]{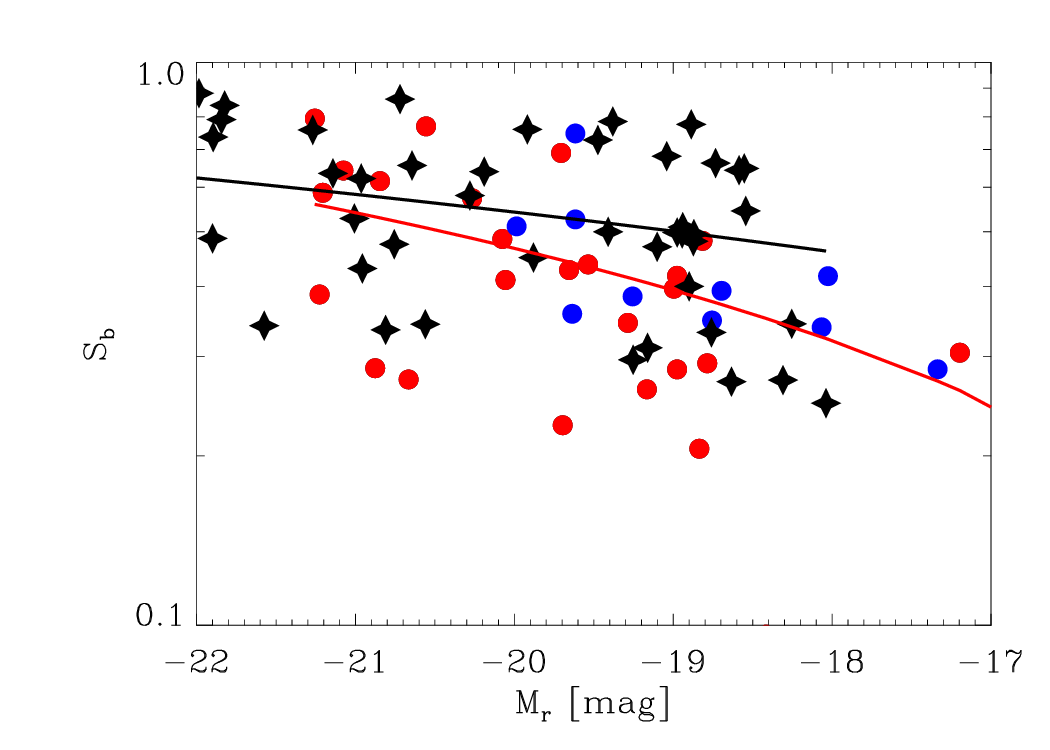}
   \caption{Bar strength for the galaxies in the Virgo cluster (blue and red points) and in the field (black stars). }
\label{sbbar}%
    \end{figure}

Figure \ref{sbbar} shows the relation between the bar strength ($S_{b}$) and the absolute magnitude of the galaxies in the cluster and in the field. The mean values of $S_{b}$ for the Virgo and the field samples are compatible within the errors (see Tab. \ref{tab:mean_value}). However, there is a weak relation between the bar strength and the absolute magnitude of the galaxies. Indeed, the slope of the relation $S_{b}-M_r$ is larger for the galaxies in the cluster ($-0.07\pm0.02$) then for the field sample ($-0.04\pm0.02$). 

Moreover, we can observe some differences in the bar strength when we divide the sample into bright ($M_{r} < -19.5$) and faint ($M_{r} > -19.5$) galaxies subsamples. Bright galaxies show similar mean values of $S_{b}$ independent of the environment ($\langle S_{\rm b}\rangle=0.51\pm0.04$ for Virgo sample and $\langle S_{\rm b}\rangle=0.59\pm0.04$ for the field galaxies). In contrast, larger differences in the mean values of $S_{b}$ can be appreciated for faint galaxies with $M_{r} > -19.5$ when comparing the cluster and field subsamples ($\langle S_{\rm b}\rangle=0.33\pm0.02$ for Virgo galaxies and $\langle S_{\rm b}\rangle=0.50\pm0.04$ for field ones). Thus, bars in the Virgo sample are weaker than in the field for faint galaxies.

No differences have been also observed in the bar strength of the bars of galaxies located in the virialized or the infall cluster regions.

\subsection{Galaxy light concentration and bar parameters}

The galaxy light concentration (C) parameter shows the fraction of light located in the innermost regions of the galaxies. This quantity have been found to correlate with galaxy properties including some from the bars \citep[see e.g. ][]{smith2022}. We have analized the light concentration of our galaxies with the bar parameters (length and strength) as a function of the environment. The galaxy light concentration was measured as: $C = 5 \times log(R_{50}/R_{90})$, being $R_{50}$ and $R_{90}$ the radii enclosing 50$\%$ and 90$\%$ of the total light of the galaxy, respectively \citep[see][]{yuho2020}.

\begin{figure}
   \centering
   \includegraphics[width=10cm]{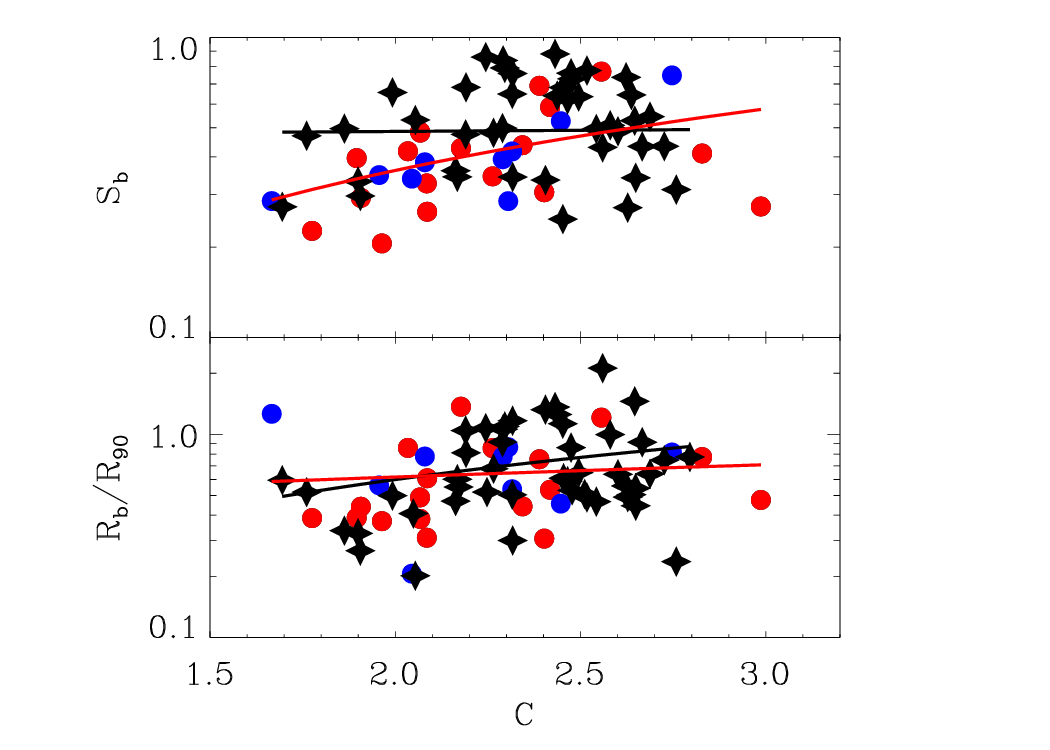}
   \caption{Galaxy light concentration as a function of bar strength (top panel) and normalized bar radius (bottom panel). The symbols represent galaxies in the Virgo cluster cluster (blue and red points) and in the field (black stars). The full lines represent the linear relations between the parameters for the galaxies in the Virgo cluster (red line) and in the field (black line). }
\label{conparam}%
    \end{figure}

Figure \ref{conparam} shows the relations between the bar strength, normalized bar radius and galaxy light concentration. In both cases the correlations are weak (less than 2$\sigma$ in the Spearman test). In particular, the bottom panel of Fig.~\ref{conparam} indicates that there is no significant differences between the normalized bar radius and the galaxy light concentration for galaxies in clusters and those in the field. In addition, galaxies with larger values of $R_{b}/R_{90}$ tend to be located in objects with larger values of $C$, regardless of the sample. The top panel of Fig.~\ref{conparam} shows that, for galaxies located in the field, there is no correlation between $S_{b}$ and $C$. In contrast, a weak correlation is observed for galaxies in the cluster. In this case, weaker bars are located in galaxies with smaller light concentration. This difference can be seen when we split the sample in two set of galaxies with $C>2.3$ and $C<2.3$. Galaxies with $C < 2.3$ show  $<S_{b}> = 0.46 \pm 0.06$ and $0.35 \pm 0.02$ for those in the field and cluster, respectively. Galaxies with $C>2.3$ show $<S_{b}> = 0.50 \pm 0.04$ and $0.49 \pm 0.05$ if they belong to the field or to the cluster, respectively. This indicates that the behaviour of the bar strength is similar regardless of galaxy environment for galaxies with high light concentration ($C>2.3$). In contrast, galaxies with small light concentrations ($C<2.3$) in the cluster show smaller values of $S_{b}$ than those located in the field. This findings suggest that a population of barred galaxies in clusters with bar strength enhanced by tidal interactions is not observed \citep[see][]{smith2022}.

\section{Discussion}

The environment plays a crucial role in the formation and evolution of galaxies. Specifically, the physical processes experienced by galaxies in clusters differ significantly from those in field environments. In high-density galaxy environments, such as galaxy clusters, tidal interactions between galaxies and the cluster potential are common \citep[see, e.g.,][]{moore1999}. These tidal interactions can range from fast fly-bys with a large impact parameter \citep[][]{moore1999} to close interactions that heavily influence the morphology and/or the stellar content of the galaxies \citep[see, e.g.,][]{gonzalezgarcia2006}. Additionally, galaxies in clusters undergo a process known as ram pressure stripping, caused by the interaction between the galaxy and the hot intracluster medium \citep[see, e.g,][]{quilis2000,Schulz2001}. These processes result in the expulsion of stars and gas from galaxies into the intracluster medium, leading to significant changes in the morphology and stellar content of galaxies in high-density environments. Consequently, late-type galaxies undergo morphological transformations, becoming early-type galaxies. This phenomenon, known as the morphology-density relation, has been observed in nearby clusters \citep{dressler1980} as well as in distant clusters \citep{cerulo2017}. The outer regions of galaxies consist of stars with lower binding energy, making them more susceptible to being stripped away during tidal interactions. Thus, these external regions are the most affected by these interactions. This effect has been observed in the discs of galaxies within clusters. Using hydro-dynamical simulations, \cite{Schulz2001} showed that gas is promptly removed from the outer disc due interaction with an hot intracluster medium, which in turn can survive from escaping immediately from the halo, even when the ram pressure stripping is not complete in the cluster cores. Moreover, \cite{Tonnesen2007} confirmed that gas is lost in galaxy cluster due to a wide variety of mechanisms, but the dominant one is a gas-only stripping event, taking place in the centre of clusters out to one viral radius and the amount of gas lost correlates with the ram-pressure the galaxy is experiencing. For instance, the disc scale of galaxies located in the Coma cluster is shorter compared to similar galaxies in the field \citep[][]{aguerri2004,gutierrez2004}. We have observed a similar effect in our galaxies, where galaxies within the Virgo cluster are shorter compared to those in the field. Furthermore, we have noticed that galaxies located in the virialized or the infall cluster region also exhibit shorter sizes compared to field galaxies. This suggests that the mechanisms responsible for the reduction in galaxy sizes are already at work during the infall process of galaxies into the cluster. This result is somehow in disagreement with the results from $N$-body simulations by \cite{lokas2016} and from cosmological hydrodynamic simulations of groups and clusters by \cite{rhee2017}, suggesting instead mass-loss due to the interaction with the cluster should be efficient after the first pericenter passage. Indeed, the infalling galaxies analysed here should not have had the time yet to reach the pericenter. On the other hand, we cannot exclude pre-processing effects could have played an important role as well.

The difference in size of galaxies with similar luminosity in clusters and field could explain the difference in the length of bars located in the Virgo cluster and the field. Specifically, bars in the field would be larger because they are located in more extended galaxies. We have seen that the difference in the bar size is not present when the bar radius is scaled by the size of the galaxy, pointing out that bars grow until reaching a fraction of the size of the disk independently of the environment in which the host galaxy is residing.

Several theoretical studies have examined the formation of bars resulting from strong tidal interactions, focusing on the effects of Milky Way (MW)-like fly-by interactions. For instance, \cite{MartinezValpuesta2016} found that fly-by encounters do not affect the radius or strength of bars when they are already formed in unstable discs. However, bars formed by interactions in stable discs tend to be shorter and weaker. In contrast, \cite{lokas2018} demonstrated that only galaxy encounters in prograde orbits, with sufficiently small impact parameters, can generate strong bar features. Nevertheless, they did not compare the structural parameters of these bars with those formed in isolation.  

The formation of bars in large-scale cosmological simulations has also been examined. In the IllustrisTNG simulation, \cite{lokas2021} investigated the properties of bars in galaxies formed through various mechanisms. The author classified barred galaxies at $z=0$ into three classes based on their interaction history. Class A comprises galaxies that experienced strong interactions resulting in significant mass loss. The bars in these galaxies formed through tidal interactions. These galaxies underwent such intense tidal effects that they exhibited bar-like structures, characterized by elongated bar-like features with small rotating discs. Class B includes galaxies in which bars formed through mergers, interactions with passing satellites, or internal disc instabilities. Class C consists of galaxies showing no evidence of strong interactions. The three classes of galaxies exhibits bars of similar strength, with a range of $A_{2, \text{max}} = 0.58-0.62$. The mean bar radii for the three classes were 4.3 kpc, 4.8 kpc, and 5.0 kpc for galaxies in classes A, B, and C, respectively. Although the differences are smaller than in their results, there is a trend towards smaller bars from field galaxies (class C) to galaxies in clusters with strong interactions (class A). However, these differences are much smaller when the bar radii are scaled. In contrast, \cite{peschken2019} found that bars in the Illustris simulation formed in galaxies experiencing strong interactions tends to be stronger than those in discs with no or minimal interactions. 

In our study, we have found that, for a given luminosity, bars formed in galaxies within the Virgo cluster are shorter than those located in the field. This contradicts the results obtained by \cite{lokas2021}, where bars are larger in galaxies experiencing strong interactions. Additionally, the overall size of galaxies, as measured by their $R_{50}$ or $R_{90}$, is also shorter in the cluster environment. However, when the radius of the bar is scaled by the size of the galaxy, the bars in the Virgo cluster and in the field appear to be similar. This suggests that larger scaled bars are found in larger galaxies. Interestingly, the growth of bars within the respective discs occurs up to a fraction of the disc size that appears to be independent of the environment. This finding implies that the process of bar formation is primarily driven by internal galactic processes rather than external ones.

Finally, we have shown a weak correlation between the bar strength and the absolute magnitude of the galaxies, which becomes remarkable when splitting the selected galaxies between bright and faint subsamples. Indeed, bars in the Virgo sample are weaker than in the field when considering the faint galaxies. A similar but weak trend was found in massive clusters as well by \cite{smith2022}. The fact that faint galaxies are weaker in the cluster than in the field could be related with the thick of their discs. Thus, discs of faint galaxies in the cluster would be thicker than those in the field producing weaker bars. This could also be in agreement with the fact that bars located in galaxies with small light concentration ($C<2.3$) show weaker bars when they are located in clusters than those in the field.

\section{Conclusion}

We have analized the properties of the bars for a total of 14 (SB0) and 22 (SBa) galaxies located in the Virgo cluster and its infall region. This galaxies span a wide luminosity range from $-22.0 < M_{r} < -17.0$. For comparison we analized a sample of 18 (SB0) and 28 (SBa) galaxies located in smaller galaxy densities environments. The radius and the strength of the bars were determined by using Fourier decomposition of the surface brightness of the $r$-band SDSS galaxy images. The main results are: 

\begin{itemize}
    \item Bars in the Virgo cluster are shorter than those located in field galaxies. In particular, the mean radius of the bars in the Virgo cluster and in the field samples are 2.6 and 6.1 kpc, respectively.
    \item There is a weaker dependence of the  strength of the bars with the environment. Thus, bright galaxies ($M_{r} < -19.5$) has similar bar strengths in the cluster and in the field. In contrast, faint galaxies ($M_{r} > -19.5$) show fainter bars in the cluster than in the field. 
    \item For a fixed luminosity, galaxies in the field are larger than those in the Virgo cluster. The mean $R_{90}$ for total sample of field and cluster galaxies are 9.3 and 3.4 kpc, respectively.
    \item The differences in the bar radius between field and Virgo galaxies are not present when they are scaled by the size of the galaxies. This indicates that independent of the environment the ratio between the bar radius and the size of the galaxy ($R_{bar}/R_{90}$) is similar for a give luminosity of the galaxies. 
\end{itemize}

The large spectroscopic galaxy surveys available today allow us, for the first time, to trace the structure of the cosmic web and analyze the properties of galaxies in environments other than galaxy clusters. In the near future, we intend to examine the characteristics of galactic bars within various settings, including clusters, filaments, and groups, in order to gain insight into the influence of the environment on bar properties.
%
  
\begin{acknowledgements}
      The research of JALA has been supported by the Spanish Ministry of Ciencia e Innovaci\'on under the grant PID2020-119342GB-I00. VC is supported by the Chilean ANID-Fondecyt Postdoctoral programme 2022 No. 3220206.
\end{acknowledgements}

\bibliographystyle{aa}
\bibliography{biblio} 

\begin{thebibliography}{66}
\expandafter\ifx\csname natexlab\endcsname\relax\def\natexlab#1{#1}\fi

\bibitem[{{Abazajian} {et~al.}(2009){Abazajian}, {Adelman-McCarthy},
  {Ag{\"u}eros}, {Allam}, {Allende Prieto}, {An}, {Anderson}, {Anderson},
  {Annis}, {Bahcall}, \& et~al.}]{Abazajian2009}
{Abazajian}, K.~N., {Adelman-McCarthy}, J.~K., {Ag{\"u}eros}, M.~A., {et~al.}
  2009, \apjs, 182, 543

\bibitem[{{Abdurro'uf} {et~al.}(2022){Abdurro'uf}, {Accetta}, {Aerts}, {Silva
  Aguirre}, {Ahumada}, {Ajgaonkar}, {Filiz Ak}, {Alam}, {Allende Prieto},
  {Almeida}, {Anders}, {Anderson}, {Andrews}, {Anguiano}, {Aquino-Ort{\'\i}z},
  {Arag{\'o}n-Salamanca}, {Argudo-Fern{\'a}ndez}, {Ata}, {Aubert},
  {Avila-Reese}, {Badenes}, {Barb{\'a}}, {Barger}, {Barrera-Ballesteros},
  {Beaton}, {Beers}, {Belfiore}, {Bender}, {Bernardi}, {Bershady}, {Beutler},
  {Bidin}, {Bird}, {Bizyaev}, {Blanc}, {Blanton}, {Boardman}, {Bolton},
  {Boquien}, {Borissova}, {Bovy}, {Brandt}, {Brown}, {Brownstein}, {Brusa},
  {Buchner}, {Bundy}, {Burchett}, {Bureau}, {Burgasser}, {Cabang}, {Campbell},
  {Cappellari}, {Carlberg}, {Wanderley}, {Carrera}, {Cash}, {Chen}, {Chen},
  {Cherinka}, {Chiappini}, {Choi}, {Chojnowski}, {Chung}, {Clerc}, {Cohen},
  {Comerford}, {Comparat}, {da Costa}, {Covey}, {Crane}, {Cruz-Gonzalez},
  {Culhane}, {Cunha}, {Dai}, {Damke}, {Darling}, {Davidson}, {Davies},
  {Dawson}, {De Lee}, {Diamond-Stanic}, {Cano-D{\'\i}az}, {S{\'a}nchez},
  {Donor}, {Duckworth}, {Dwelly}, {Eisenstein}, {Elsworth}, {Emsellem},
  {Eracleous}, {Escoffier}, {Fan}, {Farr}, {Feng}, {Fern{\'a}ndez-Trincado},
  {Feuillet}, {Filipp}, {Fillingham}, {Frinchaboy}, {Fromenteau}, {Galbany},
  {Garc{\'\i}a}, {Garc{\'\i}a-Hern{\'a}ndez}, {Ge}, {Geisler}, {Gelfand},
  {G{\'e}ron}, {Gibson}, {Goddy}, {Godoy-Rivera}, {Grabowski}, {Green},
  {Greener}, {Grier}, {Griffith}, {Guo}, {Guy}, {Hadjara}, {Harding},
  {Hasselquist}, {Hayes}, {Hearty}, {Hern{\'a}ndez}, {Hill}, {Hogg},
  {Holtzman}, {Horta}, {Hsieh}, {Hsu}, {Hsu}, {Huber}, {Huertas-Company},
  {Hutchinson}, {Hwang}, {Ibarra-Medel}, {Chitham}, {Ilha}, {Imig}, {Jaekle},
  {Jayasinghe}, {Ji}, {Johnson}, {Jones}, {J{\"o}nsson}, {Katkov}, {Khalatyan},
  {Kinemuchi}, {Kisku}, {Knapen}, {Kneib}, {Kollmeier}, {Kong}, {Kounkel},
  {Kreckel}, {Krishnarao}, {Lacerna}, {Lane}, {Langgin}, {Lavender}, {Law},
  {Lazarz}, {Leung}, {Leung}, {Lewis}, {Li}, {Li}, {Lian}, {Liang}, {Lin},
  {Lin}, {Lin}, {Lintott}, {Long}, {Longa-Pe{\~n}a}, {L{\'o}pez-Cob{\'a}},
  {Lu}, {Lundgren}, {Luo}, {Mackereth}, {de la Macorra}, {Mahadevan},
  {Majewski}, {Manchado}, {Mandeville}, {Maraston}, {Margalef-Bentabol},
  {Masseron}, {Masters}, {Mathur}, {McDermid}, {Mckay}, {Merloni},
  {Merrifield}, {Meszaros}, {Miglio}, {Di Mille}, {Minniti}, {Minsley},
  {Monachesi}, {Moon}, {Mosser}, {Mulchaey}, {Muna}, {Mu{\~n}oz}, {Myers},
  {Myers}, {Nadathur}, {Nair}, {Nandra}, {Neumann}, {Newman}, {Nidever},
  {Nikakhtar}, {Nitschelm}, {O'Connell}, {Garma-Oehmichen}, {Luan Souza de
  Oliveira}, {Olney}, {Oravetz}, {Ortigoza-Urdaneta}, {Osorio}, {Otter},
  {Pace}, {Padilla}, {Pan}, {Pan}, {Parikh}, {Parker}, {Peirani}, {Pe{\~n}a
  Ram{\'\i}rez}, {Penny}, {Percival}, {Perez-Fournon}, {Pinsonneault},
  {Poidevin}, {Poovelil}, {Price-Whelan}, {B{\'a}rbara de Andrade Queiroz},
  {Raddick}, {Ray}, {Rembold}, {Riddle}, {Riffel}, {Riffel}, {Rix}, {Robin},
  {Rodr{\'\i}guez-Puebla}, {Roman-Lopes}, {Rom{\'a}n-Z{\'u}{\~n}iga}, {Rose},
  {Ross}, {Rossi}, {Rubin}, {Salvato}, {S{\'a}nchez}, {S{\'a}nchez-Gallego},
  {Sanderson}, {Santana Rojas}, {Sarceno}, {Sarmiento}, {Sayres}, {Sazonova},
  {Schaefer}, {Schiavon}, {Schlegel}, {Schneider}, {Schultheis}, {Schwope},
  {Serenelli}, {Serna}, {Shao}, {Shapiro}, {Sharma}, {Shen}, {Shetrone}, {Shu},
  {Simon}, {Skrutskie}, {Smethurst}, {Smith}, {Sobeck}, {Spoo}, {Sprague},
  {Stark}, {Stassun}, {Steinmetz}, {Stello}, {Stone-Martinez},
  {Storchi-Bergmann}, {Stringfellow}, {Stutz}, {Su}, {Taghizadeh-Popp},
  {Talbot}, {Tayar}, {Telles}, {Teske}, {Thakar}, {Theissen}, {Tkachenko},
  {Thomas}, {Tojeiro}, {Hernandez Toledo}, {Troup}, {Trump}, {Trussler},
  {Turner}, {Tuttle}, {Unda-Sanzana}, {V{\'a}zquez-Mata}, {Valentini},
  {Valenzuela}, {Vargas-Gonz{\'a}lez}, {Vargas-Maga{\~n}a}, {Alfaro},
  {Villanova}, {Vincenzo}, {Wake}, {Warfield}, {Washington}, {Weaver},
  {Weijmans}, {Weinberg}, {Weiss}, {Westfall}, {Wild}, {Wilde}, {Wilson},
  {Wilson}, {Wilson}, {Wolf}, {Wood-Vasey}, {Yan}, {Zamora}, {Zasowski},
  {Zhang}, {Zhao}, {Zheng}, {Zheng}, \& {Zhu}}]{Abdurro'uf2022}
{Abdurro'uf}, {Accetta}, K., {Aerts}, C., {et~al.} 2022, \apjs, 259, 35

\bibitem[{{Aguerri} \& {Gonz{\'a}lez-Garc{\'\i}a}(2009)}]{Aguerri2009b}
{Aguerri}, J.~A.~L. \& {Gonz{\'a}lez-Garc{\'\i}a}, A.~C. 2009, \aap, 494, 891

\bibitem[{{Aguerri} {et~al.}(2004){Aguerri}, {Iglesias-Paramo}, {Vilchez}, \&
  {Mu{\~n}oz-Tu{\~n}{\'o}n}}]{aguerri2004}
{Aguerri}, J.~A.~L., {Iglesias-Paramo}, J., {Vilchez}, J.~M., \&
  {Mu{\~n}oz-Tu{\~n}{\'o}n}, C. 2004, \aj, 127, 1344

\bibitem[{{Aguerri} {et~al.}(2009){Aguerri}, {M{\'e}ndez-Abreu}, \&
  {Corsini}}]{Aguerri2009}
{Aguerri}, J.~A.~L., {M{\'e}ndez-Abreu}, J., \& {Corsini}, E.~M. 2009, \aap,
  495, 491

\bibitem[{{Aguerri} {et~al.}(2000){Aguerri}, {Mu{\~n}oz-Tu{\~n}{\'o}n},
  {Varela}, \& {Prieto}}]{Aguerri2000}
{Aguerri}, J.~A.~L., {Mu{\~n}oz-Tu{\~n}{\'o}n}, C., {Varela}, A.~M., \&
  {Prieto}, M. 2000, \aap, 361, 841

\bibitem[{{Athanassoula}(2003)}]{athanassoula2003}
{Athanassoula}, E. 2003, \mnras, 341, 1179

\bibitem[{{Athanassoula} {et~al.}(2013){Athanassoula}, {Machado}, \&
  {Rodionov}}]{Athanassoula2013}
{Athanassoula}, E., {Machado}, R. E.~G., \& {Rodionov}, S.~A. 2013, \mnras,
  429, 1949

\bibitem[{{Athanassoula} \& {Misiriotis}(2002)}]{Athanassoula2002}
{Athanassoula}, E. \& {Misiriotis}, A. 2002, \mnras, 330, 35

\bibitem[{{Barazza} {et~al.}(2008){Barazza}, {Jogee}, \&
  {Marinova}}]{barazza2008}
{Barazza}, F.~D., {Jogee}, S., \& {Marinova}, I. 2008, \apj, 675, 1194

\bibitem[{{Barway} {et~al.}(2011){Barway}, {Wadadekar}, \&
  {Kembhavi}}]{barway2011}
{Barway}, S., {Wadadekar}, Y., \& {Kembhavi}, A.~K. 2011, \mnras, 410, L18

\bibitem[{{Binggeli} {et~al.}(1985){Binggeli}, {Sandage}, \&
  {Tammann}}]{Binggeli1985}
{Binggeli}, B., {Sandage}, A., \& {Tammann}, G.~A. 1985, \aj, 90, 1681

\bibitem[{{Buttitta} {et~al.}(2022){Buttitta}, {Corsini}, {Cuomo}, {Aguerri},
  {Coccato}, {Costantin}, {Dalla Bont{\`a}}, {Debattista}, {Iodice},
  {M{\'e}ndez-Abreu}, {Morelli}, \& {Pizzella}}]{Buttitta2022}
{Buttitta}, C., {Corsini}, E.~M., {Cuomo}, V., {et~al.} 2022, \aap, 664, L10

\bibitem[{{Cameron} {et~al.}(2010){Cameron}, {Carollo}, {Oesch}, {Aller},
  {Bschorr}, {Cerulo}, {Aussel}, {Capak}, {Le Floc'h}, {Ilbert}, {Kneib},
  {Koekemoer}, {Leauthaud}, {Lilly}, {Massey}, {McCracken}, {Rhodes},
  {Salvato}, {Sanders}, {Scoville}, {Sheth}, {Taniguchi}, \&
  {Thompson}}]{Cameron2010}
{Cameron}, E., {Carollo}, C.~M., {Oesch}, P., {et~al.} 2010, \mnras, 409, 346

\bibitem[{{Castignani} {et~al.}(2022){Castignani}, {Vulcani}, {Finn}, {Combes},
  {Jablonka}, {Rudnick}, {Zaritsky}, {Whalen}, {Conger}, {De Lucia}, {Desai},
  {Koopmann}, {Moustakas}, {Norman}, \& {Townsend}}]{castignani2022}
{Castignani}, G., {Vulcani}, B., {Finn}, R.~A., {et~al.} 2022, \apjs, 259, 43

\bibitem[{{Cerulo} {et~al.}(2017){Cerulo}, {Couch}, {Lidman}, {Demarco},
  {Huertas-Company}, {Mei}, {S{\'a}nchez-Janssen}, {Barrientos}, \&
  {Mu{\~n}oz}}]{cerulo2017}
{Cerulo}, P., {Couch}, W.~J., {Lidman}, C., {et~al.} 2017, \mnras, 472, 254

\bibitem[{{Corsini}(2011)}]{corsini2011}
{Corsini}, E.~M. 2011, Mem. Soc. Astron. Ital. Suppl., 18, 23

\bibitem[{{Cuomo} {et~al.}(2020){Cuomo}, {Aguerri}, {Corsini}, \&
  {Debattista}}]{Cuomo2020}
{Cuomo}, V., {Aguerri}, J.~A.~L., {Corsini}, E.~M., \& {Debattista}, V.~P.
  2020, \aap, 641, A111

\bibitem[{{Cuomo} {et~al.}(2019{\natexlab{a}}){Cuomo}, {Corsini}, {Aguerri},
  {Debattista}, {Coccato}, {Costantin}, {Dalla Bont{\`a}}, {Iodice},
  {M{\'e}ndez-Abreu}, {Morelli}, {Pagotto}, \& {Pizzella}}]{Cuomo2019}
{Cuomo}, V., {Corsini}, E.~M., {Aguerri}, J.~A.~L., {et~al.}
  2019{\natexlab{a}}, \mnras, 488, 4972

\bibitem[{{Cuomo} {et~al.}(2019{\natexlab{b}}){Cuomo}, {Lopez Aguerri},
  {Corsini}, {Debattista}, {M{\'e}ndez-Abreu}, \& {Pizzella}}]{Cuomo2019b}
{Cuomo}, V., {Lopez Aguerri}, J.~A., {Corsini}, E.~M., {et~al.}
  2019{\natexlab{b}}, \aap, 632, A51

\bibitem[{{Davies} {et~al.}(2014){Davies}, {Bianchi}, {Baes}, {Bendo},
  {Clemens}, {De Looze}, {di Serego Alighieri}, {Fritz}, {Fuller},
  {Pappalardo}, {Hughes}, {Madden}, {Smith}, {Verstappen}, \&
  {Vlahakis}}]{Davies2014}
{Davies}, J.~I., {Bianchi}, S., {Baes}, M., {et~al.} 2014, \mnras, 438, 1922

\bibitem[{{Debattista} \& {Sellwood}(2000)}]{Debattista2000}
{Debattista}, V.~P. \& {Sellwood}, J.~A. 2000, \apj, 543, 704

\bibitem[{{Dressler}(1980)}]{dressler1980}
{Dressler}, A. 1980, \apj, 236, 351

\bibitem[{{Erwin} {et~al.}(2012){Erwin}, {Guti{\'e}rrez}, \&
  {Beckman}}]{Erwin2012}
{Erwin}, P., {Guti{\'e}rrez}, L., \& {Beckman}, J.~E. 2012, \apjl, 744, L11

\bibitem[{{Erwin} {et~al.}(2008){Erwin}, {Pohlen}, {Beckman}, {Guti{\'e}rrez},
  \& {Aladro}}]{Erwin2008}
{Erwin}, P., {Pohlen}, M., {Beckman}, J.~E., {Guti{\'e}rrez}, L., \& {Aladro},
  R. 2008, in Astronomical Society of the Pacific Conference Series, Vol. 390,
  Pathways Through an Eclectic Universe, ed. J.~H. {Knapen}, T.~J. {Mahoney},
  \& A.~{Vazdekis}, 251

\bibitem[{{Eskridge} {et~al.}(2000){Eskridge}, {Frogel}, {Pogge}, {Quillen},
  {Davies}, {DePoy}, {Houdashelt}, {Kuchinski}, {Ram{\'\i}rez}, {Sellgren},
  {Terndrup}, \& {Tiede}}]{eskridge2000}
{Eskridge}, P.~B., {Frogel}, J.~A., {Pogge}, R.~W., {et~al.} 2000, \aj, 119,
  536

\bibitem[{{Gnedin}(2003)}]{gnedin2003}
{Gnedin}, O.~Y. 2003, \apj, 589, 752

\bibitem[{{Gonz{\'a}lez-Garc{\'\i}a} {et~al.}(2006){Gonz{\'a}lez-Garc{\'\i}a},
  {Balcells}, \& {Olshevsky}}]{gonzalezgarcia2006}
{Gonz{\'a}lez-Garc{\'\i}a}, A.~C., {Balcells}, M., \& {Olshevsky}, V.~S. 2006,
  \mnras, 372, L78

\bibitem[{{Guo} {et~al.}(2023){Guo}, {Jogee}, {Finkelstein}, {Chen}, {Wise},
  {Bagley}, {Barro}, {Wuyts}, {Kocevski}, {Kartaltepe}, {McGrath}, {Ferguson},
  {Mobasher}, {Giavalisco}, {Lucas}, {Zavala}, {Lotz}, {Grogin},
  {Huertas-Company}, {Vega-Ferrero}, {Hathi}, {Arrabal Haro}, {Dickinson},
  {Koekemoer}, {Papovich}, {Pirzkal}, {Yung}, {Backhaus}, {Bell},
  {Calabr{\`o}}, {Cleri}, {Coogan}, {Cooper}, {Costantin}, {Croton}, {Davis},
  {Dekel}, {Franco}, {Gardner}, {Holwerda}, {Hutchison}, {Pandya},
  {P{\'e}rez-Gonz{\'a}lez}, {Ravindranath}, {Rose}, {Trump}, {de la Vega}, \&
  {Wang}}]{Guo2023}
{Guo}, Y., {Jogee}, S., {Finkelstein}, S.~L., {et~al.} 2023, \apjl, 945, L10

\bibitem[{{Guti{\'e}rrez} {et~al.}(2004){Guti{\'e}rrez}, {Trujillo}, {Aguerri},
  {Graham}, \& {Caon}}]{gutierrez2004}
{Guti{\'e}rrez}, C.~M., {Trujillo}, I., {Aguerri}, J. A.~L., {Graham}, A.~W.,
  \& {Caon}, N. 2004, \apj, 602, 664

\bibitem[{{Hoffman} {et~al.}(1980){Hoffman}, {Olson}, \&
  {Salpeter}}]{hoffman1980}
{Hoffman}, G.~L., {Olson}, D.~W., \& {Salpeter}, E.~E. 1980, \apj, 242, 861

\bibitem[{{Kashibadze} {et~al.}(2020){Kashibadze}, {Karachentsev}, \&
  {Karachentseva}}]{Kashibadze2020}
{Kashibadze}, O.~G., {Karachentsev}, I.~D., \& {Karachentseva}, V.~E. 2020,
  \aap, 635, A135

\bibitem[{{Kim} {et~al.}(2014){Kim}, {Rey}, {Jerjen}, {Lisker}, {Sung}, {Lee},
  {Chung}, {Pak}, {Yi}, \& {Lee}}]{Kim2014}
{Kim}, S., {Rey}, S.-C., {Jerjen}, H., {et~al.} 2014, \apjs, 215, 22

\bibitem[{{Li} {et~al.}(2009){Li}, {Gadotti}, {Mao}, \& {Kauffmann}}]{li2009}
{Li}, C., {Gadotti}, D.~A., {Mao}, S., \& {Kauffmann}, G. 2009, \mnras, 397,
  726

\bibitem[{{Lin} {et~al.}(2014){Lin}, {Cervantes Sodi}, {Li}, {Wang}, \&
  {Wang}}]{lin2014}
{Lin}, Y., {Cervantes Sodi}, B., {Li}, C., {Wang}, L., \& {Wang}, E. 2014,
  \apj, 796, 98

\bibitem[{{{\L}okas}(2018)}]{lokas2018}
{{\L}okas}, E.~L. 2018, \apj, 857, 6

\bibitem[{{{\L}okas}(2020)}]{Lokas2020}
{{\L}okas}, E.~L. 2020, \aap, 638, A133

\bibitem[{{{\L}okas}(2021)}]{lokas2021}
{{\L}okas}, E.~L. 2021, \aap, 647, A143

\bibitem[{{{\L}okas} {et~al.}(2016){{\L}okas}, {Ebrov{\'a}}, {del Pino},
  {Sybilska}, {Athanassoula}, {Semczuk}, {Gajda}, \& {Fouquet}}]{lokas2016}
{{\L}okas}, E.~L., {Ebrov{\'a}}, I., {del Pino}, A., {et~al.} 2016, \apj, 826,
  227

\bibitem[{{Marinova} {et~al.}(2010){Marinova}, {Jogee}, {Trentham}, {Ferguson},
  {Weinzirl}, {Balcells}, {Carter}, {den Brok}, {Erwin}, {Graham},
  {Goudfrooij}, {Guzm{\'a}n}, {Hammer}, {Hoyos}, {Peletier}, {Peng}, \&
  {Verdoes Kleijn}}]{Marinova2010}
{Marinova}, I., {Jogee}, S., {Trentham}, N., {et~al.} 2010, in Astronomical
  Society of the Pacific Conference Series, Vol. 432, New Horizons in
  Astronomy: Frank N. Bash Symposium 2009, ed. L.~M. {Stanford}, J.~D. {Green},
  L.~{Hao}, \& Y.~{Mao}, 219

\bibitem[{{Marinova} {et~al.}(2012){Marinova}, {Jogee}, {Weinzirl}, {Erwin},
  {Trentham}, {Ferguson}, {Hammer}, {den Brok}, {Graham}, {Carter}, {Balcells},
  {Goudfrooij}, {Guzm{\'a}n}, {Hoyos}, {Mobasher}, {Mouhcine}, {Peletier},
  {Peng}, \& {Verdoes Kleijn}}]{marinova2012}
{Marinova}, I., {Jogee}, S., {Weinzirl}, T., {et~al.} 2012, \apj, 746, 136

\bibitem[{{Martinez-Valpuesta} {et~al.}(2016){Martinez-Valpuesta}, {Aguerri},
  \& {Gonz{\'a}lez-Garc{\'\i}a}}]{MartinezValpuesta2016}
{Martinez-Valpuesta}, I., {Aguerri}, J., \& {Gonz{\'a}lez-Garc{\'\i}a}, C.
  2016, Galaxies, 4, 7

\bibitem[{{Martinez-Valpuesta} {et~al.}(2017){Martinez-Valpuesta}, {Aguerri},
  {Gonz{\'a}lez-Garc{\'\i}a}, {Dalla Vecchia}, \&
  {Stringer}}]{2017MNRAS.464.1502M}
{Martinez-Valpuesta}, I., {Aguerri}, J. A.~L., {Gonz{\'a}lez-Garc{\'\i}a},
  A.~C., {Dalla Vecchia}, C., \& {Stringer}, M. 2017, \mnras, 464, 1502

\bibitem[{{M{\'e}ndez-Abreu} {et~al.}(2023){M{\'e}ndez-Abreu}, {Costantin}, \&
  {Kruk}}]{MendezAbreu2023}
{M{\'e}ndez-Abreu}, J., {Costantin}, L., \& {Kruk}, S. 2023, arXiv e-prints,
  arXiv:2307.02898

\bibitem[{{M{\'e}ndez-Abreu} {et~al.}(2017){M{\'e}ndez-Abreu}, {Ruiz-Lara},
  {S{\'a}nchez-Menguiano}, {de Lorenzo-C{\'a}ceres}, {Costantin},
  {Catal{\'a}n-Torrecilla}, {Florido}, {Aguerri}, {Bland-Hawthorn}, {Corsini},
  {Dettmar}, {Galbany}, {Garc{\'{\i}}a-Benito}, {Marino}, {M{\'a}rquez},
  {Ortega-Minakata}, {Papaderos}, {S{\'a}nchez}, {S{\'a}nchez-Blazquez},
  {Spekkens}, {van de Ven}, {Wild}, \& {Ziegler}}]{MendezAbreu2017}
{M{\'e}ndez-Abreu}, J., {Ruiz-Lara}, T., {S{\'a}nchez-Menguiano}, L., {et~al.}
  2017, \aap, 598, A32

\bibitem[{{M{\'e}ndez-Abreu} {et~al.}(2010){M{\'e}ndez-Abreu},
  {S{\'a}nchez-Janssen}, \& {Aguerri}}]{MendezAbreu2010}
{M{\'e}ndez-Abreu}, J., {S{\'a}nchez-Janssen}, R., \& {Aguerri}, J.~A.~L. 2010,
  \apjl, 711, L61

\bibitem[{{M{\'e}ndez-Abreu} {et~al.}(2012){M{\'e}ndez-Abreu},
  {S{\'a}nchez-Janssen}, {Aguerri}, {Corsini}, \&
  {Zarattini}}]{MendezAbreu2012}
{M{\'e}ndez-Abreu}, J., {S{\'a}nchez-Janssen}, R., {Aguerri}, J.~A.~L.,
  {Corsini}, E.~M., \& {Zarattini}, S. 2012, \apjl, 761, L6

\bibitem[{{Moore} {et~al.}(1999){Moore}, {Lake}, {Quinn}, \&
  {Stadel}}]{moore1999}
{Moore}, B., {Lake}, G., {Quinn}, T., \& {Stadel}, J. 1999, \mnras, 304, 465

\bibitem[{{Ohta} {et~al.}(1990){Ohta}, {Hamabe}, \& {Wakamatsu}}]{Ohta1990}
{Ohta}, K., {Hamabe}, M., \& {Wakamatsu}, K.-I. 1990, \apj, 357, 71

\bibitem[{{Oman} {et~al.}(2013){Oman}, {Hudson}, \& {Behroozi}}]{oman2013}
{Oman}, K.~A., {Hudson}, M.~J., \& {Behroozi}, P.~S. 2013, \mnras, 431, 2307

\bibitem[{{Peschken} \& {{\L}okas}(2019)}]{peschken2019}
{Peschken}, N. \& {{\L}okas}, E.~L. 2019, \mnras, 483, 2721

\bibitem[{{Quilis} {et~al.}(2000){Quilis}, {Moore}, \& {Bower}}]{quilis2000}
{Quilis}, V., {Moore}, B., \& {Bower}, R. 2000, Science, 288, 1617

\bibitem[{{Raha} {et~al.}(1991){Raha}, {Sellwood}, {James}, \&
  {Kahn}}]{Raha1991}
{Raha}, N., {Sellwood}, J.~A., {James}, R.~A., \& {Kahn}, F.~D. 1991, \nat,
  352, 411

\bibitem[{{Rhee} {et~al.}(2017){Rhee}, {Smith}, {Choi}, {Yi}, {Jaff{\'e}},
  {Candlish}, \& {S{\'a}nchez-J{\'a}nssen}}]{rhee2017}
{Rhee}, J., {Smith}, R., {Choi}, H., {et~al.} 2017, \apj, 843, 128

\bibitem[{{S{\'a}nchez} {et~al.}(2012){S{\'a}nchez}, {Kennicutt}, {Gil de Paz},
  {van de Ven}, {V{\'\i}lchez}, {Wisotzki}, {Walcher}, {Mast}, {Aguerri},
  {Albiol-P{\'e}rez}, {Alonso-Herrero}, {Alves}, {Bakos}, {Bart{\'a}kov{\'a}},
  {Bland-Hawthorn}, {Boselli}, {Bomans}, {Castillo-Morales}, {Cortijo-Ferrero},
  {de Lorenzo-C{\'a}ceres}, {Del Olmo}, {Dettmar}, {D{\'\i}az}, {Ellis},
  {Falc{\'o}n-Barroso}, {Flores}, {Gallazzi}, {Garc{\'\i}a-Lorenzo},
  {Gonz{\'a}lez Delgado}, {Gruel}, {Haines}, {Hao}, {Husemann},
  {Igl{\'e}sias-P{\'a}ramo}, {Jahnke}, {Johnson}, {Jungwiert}, {Kalinova},
  {Kehrig}, {Kupko}, {L{\'o}pez-S{\'a}nchez}, {Lyubenova}, {Marino},
  {M{\'a}rmol-Queralt{\'o}}, {M{\'a}rquez}, {Masegosa}, {Meidt},
  {Mendez-Abreu}, {Monreal-Ibero}, {Montijo}, {Mour{\~a}o}, {Palacios-Navarro},
  {Papaderos}, {Pasquali}, {Peletier}, {P{\'e}rez}, {P{\'e}rez}, {Quirrenbach},
  {Rela{\~n}o}, {Rosales-Ortega}, {Roth}, {Ruiz-Lara},
  {S{\'a}nchez-Bl{\'a}zquez}, {Sengupta}, {Singh}, {Stanishev}, {Trager},
  {Vazdekis}, {Viironen}, {Wild}, {Zibetti}, \& {Ziegler}}]{Sanchez2012}
{S{\'a}nchez}, S.~F., {Kennicutt}, R.~C., {Gil de Paz}, A., {et~al.} 2012,
  \aap, 538, A8

\bibitem[{{Schulz} \& {Struck}(2001)}]{Schulz2001}
{Schulz}, S. \& {Struck}, C. 2001, \mnras, 328, 185

\bibitem[{{Skibba} {et~al.}(2012){Skibba}, {Masters}, {Nichol}, {Zehavi},
  {Hoyle}, {Edmondson}, {Bamford}, {Cardamone}, {Keel}, {Lintott}, \&
  {Schawinski}}]{skibba2012}
{Skibba}, R.~A., {Masters}, K.~L., {Nichol}, R.~C., {et~al.} 2012, \mnras, 423,
  1485

\bibitem[{{Smith} {et~al.}(2022){Smith}, {Giroux}, \& {Struck}}]{smith2022}
{Smith}, B.~J., {Giroux}, M.~L., \& {Struck}, C. 2022, \aj, 164, 146

\bibitem[{{Smith} {et~al.}(2015){Smith}, {S{\'a}nchez-Janssen}, {Beasley},
  {Candlish}, {Gibson}, {Puzia}, {Janz}, {Knebe}, {Aguerri}, {Lisker},
  {Hensler}, {Fellhauer}, {Ferrarese}, \& {Yi}}]{smith2015}
{Smith}, R., {S{\'a}nchez-Janssen}, R., {Beasley}, M.~A., {et~al.} 2015,
  \mnras, 454, 2502

\bibitem[{{Tawfeek} {et~al.}(2022){Tawfeek}, {Cervantes Sodi}, {Fritz},
  {Moretti}, {P{\'e}rez-Mill{\'a}n}, {Gullieuszik}, {Poggianti}, {Vulcani}, \&
  {Bettoni}}]{Tawfeek2022}
{Tawfeek}, A.~A., {Cervantes Sodi}, B., {Fritz}, J., {et~al.} 2022, \apj, 940,
  1

\bibitem[{{Thompson}(1981)}]{thompson1981}
{Thompson}, L.~A. 1981, \apjl, 244, L43

\bibitem[{{Tonnesen} {et~al.}(2007){Tonnesen}, {Bryan}, \& {van
  Gorkom}}]{Tonnesen2007}
{Tonnesen}, S., {Bryan}, G.~L., \& {van Gorkom}, J.~H. 2007, \apj, 671, 1434

\bibitem[{{van den Bergh}(2002)}]{vandenbergh2002}
{van den Bergh}, S. 2002, \aj, 124, 782

\bibitem[{{Walcher} {et~al.}(2014){Walcher}, {Wisotzki}, {Bekerait{\'e}},
  {Husemann}, {Iglesias-P{\'a}ramo}, {Backsmann}, {Barrera Ballesteros},
  {Catal{\'a}n-Torrecilla}, {Cortijo}, {del Olmo}, {Garcia Lorenzo},
  {Falc{\'o}n-Barroso}, {Jilkova}, {Kalinova}, {Mast}, {Marino},
  {M{\'e}ndez-Abreu}, {Pasquali}, {S{\'a}nchez}, {Trager}, {Zibetti},
  {Aguerri}, {Alves}, {Bland-Hawthorn}, {Boselli}, {Castillo Morales}, {Cid
  Fernandes}, {Flores}, {Galbany}, {Gallazzi}, {Garc{\'{\i}}a-Benito}, {Gil de
  Paz}, {Gonz{\'a}lez-Delgado}, {Jahnke}, {Jungwiert}, {Kehrig}, {Lyubenova},
  {M{\'a}rquez Perez}, {Masegosa}, {Monreal Ibero}, {P{\'e}rez}, {Quirrenbach},
  {Rosales-Ortega}, {Roth}, {Sanchez-Blazquez}, {Spekkens}, {Tundo}, {van de
  Ven}, {Verheijen}, {Vilchez}, \& {Ziegler}}]{Walcher2014}
{Walcher}, C.~J., {Wisotzki}, L., {Bekerait{\'e}}, S., {et~al.} 2014, \aap,
  569, A1

\bibitem[{{Yoon} {et~al.}(2019){Yoon}, {Im}, {Lee}, {Lee}, \& {Lim}}]{Yoon2019}
{Yoon}, Y., {Im}, M., {Lee}, G.-H., {Lee}, S.-K., \& {Lim}, G. 2019, Nature
  Astronomy, 3, 844

\bibitem[{{Yu} \& {Ho}(2020)}]{yuho2020}
{Yu}, S.-Y. \& {Ho}, L.~C. 2020, \apj, 900, 150

\end{thebibliography}

\end{document}